\documentstyle[preprint,aps]{revtex}

\begin{document}
\draft
\title{A Wigner-Seitz model of charged lamellar colloidal dispersions.}
\author{Emmanuel Trizac 
\footnote{E-mail: etrizac@physique.ens-lyon.fr}
and Jean-Pierre Hansen \footnote{E-mail: hansen@physique.ens-lyon.fr}
}
\address{Laboratoire de Physique, URA 1325 du CNRS; \\
Ecole Normale Sup\'erieure de Lyon; 69364 Lyon Cedex 07 (France)
}

\maketitle

\begin{abstract}
A concentrated suspension of lamellar colloidal particles (e. g. 
clay) is modelled by considering a single, uniformly charged,
finite platelet confined with co- and counterions to a Wigner-Seitz (WS)
cell. The system is treated within Poisson-Boltzmann theory,
with appropriate boundary conditions on the surface of the WS cell,
supposed to account for the confinement effect of neighbouring 
platelets. Expressions are obtained for the free energy,
osmotic and disjoining pressures and the capacitance in terms
of the local electrostatic potential and the co- and counterion
density profiles. Explicit solutions of the {\em linearized}
Poisson-Boltzmann (LPB) equation are obtained for circular
and square platelets placed at the centre of a cylindrical
or parallelepipedic cell. The resulting free energy
is found to go through a minimum as a function of
the aspect ratio of the cell, for any given volume (determined
by the macroscopic concentration of platelets), platelet surface charge and
salt concentration. The optimum aspect ratio is found
to be nearly independent of the two latter physical parameters.
The osmotic and disjoining pressures are found to coincide at the free
energy minimum, while the total quadrupole moment of the 
electric double-layer formed by the platelet and the surrounding 
co- and counterions vanishes simultaneously. The osmotic
equation-of-state is calculated for a variety of physical conditions.
The limit of vanishing platelet concentration is considered
in some detail, and the force acting between two 
coaxial platelets is calculated in that limit
as a function of their separation.

\end{abstract}


\bigskip

\newpage

\narrowtext

\section{Introduction}
\label{sec:intro}
Charge-stabilized colloidal suspensions have been the object of intense 
theoretical scrutiny, starting with the classic work of Derjaguin and Landau,
and of Verwey and Overbeek (DLVO) \cite{VeOv}, on interacting electric 
double-layers, which took its roots in the even earlier work of Gouy 
\cite{Gouy} 
and Chapman \cite{Chap} on double-layers near infinite, uniformly
charged planes. Following the latter pioneering contributions, a large body 
of work has been devoted to the planar geometry, which is relevant for electric
double-layers near macroscopic electrodes \cite{CaTo}, or near membranes
of biological interest \cite{Ande}. DLVO theory applies
to spherical geometry in particular, 
and yields an effective interaction between 
the electric double-layers surrounding spherical colloidal particles,
in the form of a screened Coulomb potential, the validity of which
has been tested both theoretically \cite{LoHM,LoKr} and experimentally
\cite{CrGr}. The cylindrical geometry, appropriate for (infinitely)
long, stiff polyelectrolyte chains, has been investigated along similar
lines, starting with the work of Katchalsky and collaborators
\cite{FuKL,BaJo}. The common framework of much of this work is
Poisson-Boltzmann (PB) theory, which is a mean field approximation 
within the density functional theory of inhomogeneous fluids \cite{EvHS}.

Poisson-Boltzmann theory applies equally well to the description 
of the inhomogeneous distributions of co- and counterions around
an isolated charged colloidal particle (or polyion) suspended 
in an ionic solution, and to a concentrated suspension of polyions, 
provided the ``cage'' of neighbouring polyions is modelled by a 
Wigner-Seitz (WS) cell of appropriate geometry, to which each polyion
is confined. Given proper boundary conditions, the Wigner-Seitz
model reduces the initial many-polyion suspension to the much simpler 
problem of a single polyion confined with its associated co- and counterions
to a cell of volume equal to the volume per polyion of the dispersion. 
The shape of the WS cell should reflect the shape of the polyion: it will
be an infinite slab for a membrane \cite{Ande}, an infinite coaxial cylinder
in the case of a linear polyelectrolyte \cite{FuKL}, or a concentric
sphere surrounding a spherical colloid \cite{ACGMPH}.

Except in the case of spheres, the aforementioned models and calculations
deal with the case of polyions of infinite extension ({\it e. g.} an
infinite plane in the case of membranes, or an infinite line or cylinder
in the case of polyelectrolyte chains). In this paper, we examine the case
of rigid membranes or platelets of finite size. Restriction
will be made to infinitely thin, uniformly charged circular or square
platelets. These may be considered as a reasonable model for dispersions 
of smectite clay particles, like the natural montmorillonite
clays \cite{vOlp} or the synthetic Laponite clays \cite{LILT}.
While the former are irregularly shaped polygonal particles,
the latter are, to a good approximation, of a circular shape
(disc-like particles). The thickness of a real clay particle is 
of the order of 1~nm, which is much less than lateral dimensions
(Laponite discs have a diameter of typically 25 nm), so that the picture
of infinitely thin platelets is acceptable. The main difficulty in
a statistical description of such platelets lies in the considerable
anisotropy of the particles and of their associated electric 
double-layers, compared to the much-studied case of spherical colloids.  

The main objective of this paper is to obtain expressions for the 
density profiles of co- and counterions around a circular or square
platelet of finite size, confined to WS cells of cylindrical or 
parallelepipedic geometry. The calculations are carried
out within linearized Poisson-Boltzmann theory, and the calculated 
profiles and potential distributions will be used to evaluate 
key equilibrium properties of concentrated lamellar suspensions, including
the free energy, the stress tensor, the capacitance and the quadrupole moment
of a platelet and its associated electric double-layer.

Preliminary accounts of parts of this work have appeared elsewhere
\cite{HaTr,TrHa1,TrHa2}.

\section{Density functional and Poisson-Boltzmann theory}
\label{sec:dft}
It is instructive to repeat the derivation of the familiar Poisson-Boltzmann
equation from the free energy functional of the inhomogeneous fluid
of co- and counterions contained in a WS cell, and subjected to the
``external'' electric field due to the uniform charge density on the finite
platelet placed at the centre of the WS cell. In the case of smectite clays,
the surface charge density $\sigma = -Ze/\Sigma_p$ (where $\Sigma_p$
is the area of the platelet and $-Ze$ the total charge, in multiples 
of the proton charge $e>0$) is negative, so that the counterions are positive,
while the co-ions are negative; both are assumed to be monovalent. The local 
co- and counterion densities (or density profiles) are denoted by 
$\rho^{-}({\bf r})$ and $\rho^{+}({\bf r})$, where ${\bf r}$ is a position
vector pointing inside the WS cell. If $N^{+}$ and $N^{-}$ are the total
numbers of co- and counterions inside the WS cell, the $\rho^{\alpha}({\bf r})$
satisfy the normalization condition:
\begin{equation}
\int_{\Omega} \rho^{\alpha}({\bf r})\, d^3{\bf r} = N^{\alpha}~; \quad
\alpha = +, -
\label{eq:normal}
\end{equation}
where $\Omega$ is the volume of the WS cell, 
which is equal to the average volume per platelet in the colloidal
suspension. Overall charge neutrality requires that:
\begin{equation}
N^{+}-N^{-} -Z = 0.
\label{eq:neutra}
\end{equation}
The free energy functional ${\cal F}[\rho^{+},\rho^{-}]$ may be split
into the usual ideal, Coulombic, external and correlational contributions
\cite{EvHS}:
\begin{equation}
{\cal F} = 
{\cal F}_{\hbox{\scriptsize id}} +  
{\cal F}_{\hbox{\scriptsize Coul}} + 
{\cal F}_{\hbox{\scriptsize ext}} + 
{\cal F}_{\hbox{\scriptsize corr}} 
\label{eq:dft_gen}
\end{equation}
with
\begin{eqnarray*}
&&{\cal F}_{\hbox{\scriptsize id}}[\{\rho^{\alpha}\}] = 
k_{_{B}} T \sum_{\alpha = +,-} \int_{\Omega} 
d^3{\bf r}~\rho^{\alpha}({\bf r})~[\,\ln(\Lambda_{\alpha}^3\,
\rho^{\alpha}({\bf r})) - 1\,] 
\nonumber \\
&&{\cal F}_{\hbox{\scriptsize Coul}}[\{\rho^{\alpha}\}] = 
\frac{e^2}{2\varepsilon} \int_{\Omega} d^3{\bf r}\int_{\Omega} d^3{\bf r}'
 ~\frac{\rho_c({\bf r}) \, \rho_c({\bf r}')}{|{\bf r}-{\bf r}'|}
\nonumber \\
&&{\cal F}_{\hbox{\scriptsize ext}}[\{\rho^{\alpha}\}]= \frac{e}{\varepsilon} 
\int_{\Omega} d^3{\bf r}~\rho_c({\bf r})~\varphi_{_{\cal P}}({\bf r}). 
\nonumber \\
\end{eqnarray*}
In these equations $\varepsilon$ is the dielectric constant of the solvent
(generally water) regarded as a continuous medium (``primitive model'');
$\Lambda_{\alpha}$ is the de Broglie thermal wavelength of ions 
of species $\alpha$; $\rho_c({\bf r})=\rho^+({\bf r})-\rho^-({\bf r})$
is the local charge density (in unit of $e$); and 
$\varphi_{_{\cal P}}({\bf r})$ is the electrostatic potential at ${\bf r}$ 
due to the uniformly charged platelet, the centre of which coincides 
with the centre of the WS cell, conveniently chosen as the origin.
The correlation part, ${\cal F}_{\hbox{\scriptsize corr}}$, is not known
explicitly, but may be expressed within the local density approximation
\cite{EvHS,LoHM}. ${\cal F}_{\hbox{\scriptsize corr}}$ will be neglected
throughout, an approximation which is reasonable, as long as the local 
concentrations of co- and counterions are not too high, a condition 
which may not be fulfilled for the counterions in the immediate
vicinity of a highly
charged platelet. For a complete formulation of the electrostatic
problem, the form (\ref{eq:dft_gen}) of the free energy functional must
be supplemented by specifying the boundary conditions satisfied
by the resulting mean electrostatic potential, or its gradient,
on the surface $\Sigma$ of the WS cell.

The equilibrium density profiles $\rho^{\alpha}({\bf r})$ are those
which minimize the free energy functional (\ref{eq:dft_gen}), 
subject to the constraints (\ref{eq:normal}) {\it i. e.} 
\begin{equation}
\frac{\delta {\cal F}[\rho^{+},\rho^{-}]}{\delta \rho^{\alpha}({\bf r})}
= \mu_{\alpha}~; \quad \alpha = +, -
\label{eq:variat}
\end{equation}
where $\mu_{\alpha}$ is the Lagrange multiplier associated with the constraint
(\ref{eq:normal}), {\it i. e.} the chemical potential of species $\alpha$. 
With ${\cal F}_{\hbox{\scriptsize corr}}=0$, the functional derivatives 
are easily calculated and the optimum density profiles are found
to be given by the Boltzmann distribution:
\begin{equation}
\rho^\pm({\bf r}) = \rho_{_{0}}^\pm\exp\left\{\mp\beta e \, 
\varphi({\bf r})\right\}
\label{eq:boltzm}
\end{equation}
where $\beta=1/(k_{_{B}}T)$ is the inverse temperature in energy units,
and $\varphi({\bf r})$ is the total electrostatic potential at ${\bf r}$:
\begin{equation}
\varphi({\bf r}) = \varphi_{_{\cal P}} ({\bf r}) + \frac{e}{\varepsilon}
\int_{\Omega} d^3{\bf r}'\, \frac{\rho_c({\bf r}')}{|{\bf r}-{\bf r}'|}
\label{eq:potgen}
\end{equation}
which satisfies Poisson's equation:
\begin{equation}
\nabla^2 \varphi({\bf r}) = -\frac{4\pi}{\varepsilon} \, 
q_{_{\cal P}}({\bf r})  -\frac{4\pi e}{\varepsilon} \, 
\rho_c({\bf r}).
\label{eq:poisso}
\end{equation}
In eq. (\ref{eq:poisso}), $q_{_{\cal P}}({\bf r})$ denotes the
(surface) charge density of the platelet. For a given 
$q_{_{\cal P}}({\bf r})$, eqs. (\ref{eq:boltzm}) and 
(\ref{eq:poisso}) form a closed set, which may be reexpressed as an 
inhomogeneous nonlinear partial differential equation for the potential
$\varphi({\bf r})$, usually referred to as the Poisson-Boltzmann
(PB) equation:
\begin{equation}
\nabla^2 \varphi({\bf r}) + \frac{4\pi e}{\varepsilon}\,\left[
\rho_{_{0}}^+\exp\left\{-\beta e \, \varphi({\bf r})\right\}-
\rho_{_{0}}^-\exp\left\{+\beta e \, \varphi({\bf r})\right\} \right]
= -\frac{4\pi}{\varepsilon}\,q_{_{\cal P}}({\bf r}).
\label{eq:pb}
\end{equation}
Neglect of ${\cal F}_{\hbox{\scriptsize corr}}$ clearly points to 
the mean-field nature of PB theory. The prefactors $\rho_{_{0}}^\pm$
are equal to the fugacities $\exp(\beta\mu_\pm) /\Lambda_{\pm}$ in the
case of an open system, where the WS cell is in equilibrium 
with an infinite reservoir which fixes the chemical potentials 
$\mu_\pm$. For a suspension of fixed ionic composition (canonical ensemble),
the $\rho_{_{0}}^\pm$ are determined from the normalization conditions 
(\ref{eq:normal}). Equation (\ref{eq:pb}) must be solved, subject
to appropriate boundary conditions on the surface $\Sigma$ of the WS cell.
If the surface is regarded as a boundary between WS cells associated 
with nearest neighbour platelets, it is natural to impose that the
{\em normal} component of the electric field ${\bf E} = -
\bbox{\nabla} \varphi$
vanish at each point on $\Sigma$. In practice, the homogeneous version
of the PB equation (\ref{eq:pb}) is solved for all positions 
${\bf r}$ outside the platelet (${\bf r} \notin \Sigma_p$), and the usual
discontinuity of the normal component of the field upon crossing 
the uniformly charged surface $\Sigma_p$ is taken into account.

The PB equation has been solved numerically for circular platelets
of finite thickness ({\it i. e.} coin-like cylinders), carrying a 
positive edge charge, in the limit of an infinitely dilute suspension
$(\Omega \to \infty)$ \cite{SRCS}. In this case, the boundary 
$\Sigma$ of the WS cell is pushed out to infinity, the electrostatic potential 
$\varphi({\bf r})$ can be chosen to vanish when $|{\bf r}| \to \infty$, and
the prefactors $\rho_{_{0}}^\pm$ reduce to the macroscopic co- and 
counterion number concentrations $n^-$ and $n^+$ ($n^+=n^-$). An analytic
solution of the PB equation (\ref{eq:pb}), for vanishing or finite 
platelet concentration $n = 1/\Omega$, is available only
in the limit of an infinite platelet in a WS slab ; this geometry reduces 
the problem to the classic one-dimensional Gouy-Chapman problem
\cite{Gouy,Chap}. For finite platelets, analytic solutions can be obtained 
only upon linearization of the PB equation (\ref{eq:pb}). 
For this purpose, it is convenient to redefine the prefactors 
$\rho_{_{0}}^\pm$ such that
\begin{equation}
\rho^\pm({\bf r}) = \rho_{_{0}}^\pm\exp\left\{\mp\beta e \, 
\left[\varphi({\bf r})-\varphi^\ast\right]\right\}
\label{eq:phista}
\end{equation}
where $\varphi^\ast$ is a reference potential to be specified. The resulting
linearized PB equation reads:
\begin{equation}
\nabla^{2} \varphi({\bf r}) -
\kappa_{_{D}}^{2} \left[\, \varphi({\bf r}) -\gamma_{\scriptscriptstyle 0}\, \right] =
-\frac{4 \pi}{\varepsilon}\, q_{_{\cal P}}({\bf r}) 
\label{eq:dh}
\end{equation}
where the squared inverse Debye length $\kappa_{_{D}}^{2}=1/\lambda_{_{D}}^2$,
and the constant $\gamma_{\scriptscriptstyle 0}$ are given by:
\begin{mathletters}
\begin{eqnarray}
&&\kappa_{_{D}}^{2} = 4\pi \ell_{_{B}}\left(\rho_{_{0}}^++\rho_{_{0}}^-
\right)\\
&&\gamma_{\scriptscriptstyle 0} = 
\left(\rho_{_{0}}^+-\rho_{_{0}}^-\right)  \frac{4 \pi e}{
\varepsilon \kappa_{_{D}}^2} \,+ \,\varphi^\ast\, =\, 
\frac{kT}{e}\,\frac{\rho_{_{0}}^+-\rho_{_{0}}^-}
{\rho_{_{0}}^++\rho_{_{ 0}}^-}\,+ \,\varphi^\ast
\end{eqnarray}
\end{mathletters}
\noindent and $\ell_{_{B}} = \beta e^2/\varepsilon$ is the Bjerrum length
($\ell_{_{B}} \simeq 0.7$ nm in water at room temperature). 
The normalization conditions (\ref{eq:normal}) now reduce to:
\begin{equation}
n^\pm = \frac{N^\pm}{\Omega} = \rho_{_{0}}^\pm\, \left[1\pm \beta e 
\left(\varphi^\ast-\overline{\varphi}\right)\right]
\label{eq:normdh}
\end{equation}
where $\overline{\varphi}$ is the mean potential in the WS cell:
\begin{equation}
\overline{\varphi} = \frac{1}{\Omega} \, \int_{\Omega} \, 
\varphi({\bf r}) \, d^3{\bf r}.
\label{eq:phibarre}
\end{equation}
A particularly simple choice is thus to linearize the local densities
around $\overline{\varphi}$, {\it i. e.} to take $\varphi^\ast
=\overline{\varphi}$ \cite{TrHa1}, in which case
\begin{equation}
\rho_{_{0}}^\pm = n^\pm \qquad \hbox{and} \qquad 
\kappa_{_{D}}^{2} = 4\pi \ell_{_{B}}\left(n^++n^-\right).
\end{equation}
It is worthwhile to note that linearized Poisson-Boltzmann (LPB) theory
may be derived from the free energy functional (\ref{eq:dft_gen})
(with ${\cal F}_{\hbox{\scriptsize corr}}=0$) via the variational principle
(\ref{eq:variat}), provided the integrand in the ``ideal'' contribution
${\cal F}_{\hbox{\scriptsize id}}$, is expanded to second order in powers of 
the local densities $\rho^\alpha({\bf r})$ from their mean $n^\alpha$
\cite{LoHM}. 

Before turning to the presentation of explicit solutions of LPB theory
for specific geometries, we address the problem of expressing and calculating 
key macroscopic quantities like the free energy, the stress tensor and
the osmotic pressure, from the density profiles.

\section{Free energy and pressure tensor}
\label{sec:free}

The Helmholz free energy $F$ of the electric double-layer
around a colloidal particle inside a WS cell is the key thermodynamic
quantity which must be evaluated with care \cite{VeOv,Ande,Marc}. 
Within mean-field PB theory, the free energy $F$ may, in principle,
be calculated by substituting the equilibrium density profiles
$\rho^{\alpha}({\bf r})$, determined via the variational principle
(\ref{eq:variat}), into the functional (\ref{eq:dft_gen}) (with
${\cal F}_{\hbox{\scriptsize corr}}=0$). This expression of 
the free energy may be cast in the standard form
\begin{equation}
F = U- T S,
\label{eq:futs}
\end{equation}
where the internal energy and the entropy are given by \cite{TrHa1}
\begin{eqnarray}
U &=& \frac{1}{2}\,\int_{\Omega}\left\{q_{_{\cal P}}({\bf r})
+ e\left[\rho^+({\bf r})-\rho^-({\bf r})\right]\right\}\, \varphi({\bf r})
\, d^3{\bf r} 
\nonumber \\
&=& \frac{\varepsilon}{8\pi}\, \int_{\Omega} \left[\bbox{\nabla}
\varphi({\bf r})\right]^2\, d^3{\bf r} \,- \,\frac{\varepsilon}{8\pi}
\oint_{\Sigma} \varphi({\bf r}) \bbox{\nabla}\varphi({\bf r}) \cdot 
\bbox{d S}
\label{eq:u}
\end{eqnarray}
and
\begin{eqnarray}
T S =  - k_{_{B}}T &&\sum_{\alpha = +,-} \int_{\Omega} 
\rho^{\alpha}({\bf r})~[\,\ln(\Lambda_{\alpha}^3\,
\rho^{\alpha}({\bf r})) - 1\,]\,d^3{\bf r}
\nonumber \\
= - k_{_{B}}T&& \sum_{\alpha = +,-}  N^\alpha
\ln\left(\rho_{_{0}}^\alpha \Lambda_\alpha^3  \right) + 
\nonumber\\
&&\int_{\Omega}\left\{e\left[\rho^+({\bf r})-\rho^-({\bf r})\right]
\varphi({\bf r}) + k_{_{b}}T\left[\rho^+({\bf r})+\rho^-({\bf r})\right]\right\}
\, d^3{\bf r}.
\label{eq:s}
\end{eqnarray}
In eqs. (\ref{eq:u}) and (\ref{eq:s}), the $\rho^{\alpha}({\bf r})$
are the equilibrium profiles, and use was made of eqs. (\ref{eq:poisso})
and (\ref{eq:boltzm}) in going from the first to the second line in each
of these equations. The resulting expression for the dimensionless
free energy $\beta F$ reads, in terms of the dimensionless
electrostatic potential $\Phi({\bf r}) = \beta e \varphi({\bf r})$:
\begin{eqnarray}
\beta F &&=  \sum_{\alpha = +,-}  N^\alpha
\ln\left(\rho_{_{0}}^\alpha \Lambda_\alpha^3  \right)
 -\frac{1}{8\pi\ell_{_{B}}} 
\oint_{\Sigma} \Phi({\bf r}) \,\bbox{\nabla}\, \Phi({\bf r})
\cdot \bbox{d S} \nonumber\\
&&
 +\int_{\Omega} \left\{
\left[\rho^-({\bf r})-\rho^+({\bf r})\right]\Phi({\bf r}) - 
\left[\rho^+({\bf r}) + \rho^-({\bf r})\right]
+ \frac{1}{8\pi\ell_{_{B}}} 
\left[\bbox{\nabla} \Phi({\bf r})\right]^2 \right\}\,d^3{\bf r}.
\label{eq:freepb}
\end{eqnarray}
Expression (\ref{eq:freepb}), valid within the non-linear 
PB approximation, involves integrations over the WS boundary surface
(which do not contribute if the boundary condition of vanishing normal
electric field is adopted) and over the volume of the WS cell, and
is hence not very tractable. Equivalent expressions for the free
energy can be obtained by considering generic charging processes
\cite{VeOv,Ande,TrHa1,Marc}. For a fixed cell geometry (volume and shape),
the variation in free energy due to infinitesimal variations 
of the potential $(\Phi({\bf r}) \to \Phi({\bf r}) + \delta \Phi({\bf r}))$
and of the Bjerrum length ($\ell_{_{B}} \to \ell_{_{B}}+\delta \ell_{_{B}}$),
is of the generic form \cite{TrHa1}
\begin{eqnarray}
\delta\left(\beta F\right) =&& \frac{1}{8\pi \ell_{_{B}}} 
\oint_{\Sigma} \left[\Phi 
\,\bbox{\nabla} \left(\delta \Phi \right) - \delta \Phi 
\,\bbox{\nabla} \Phi \right] \cdot \bbox{d S} + 
\beta U \,\frac{\delta \ell_{_{B}}}{\ell_{_{B}}} \nonumber\\
&& + \int_{\Sigma_p} \Phi \, \delta \left(\frac{\sigma}{e}\right) \,  d^2{\bf r}
\,+\, \left[\ln\left(\rho_{_{0}}^+\Lambda_+^3\right) \right]\delta N_+ \,+\,
 \left[\ln\left(\rho_{_{0}}^-\Lambda_-^3\right) \right]\delta N_-,
\label{eq:deltaF}
\end{eqnarray}
where $\sigma = -Ze/\Sigma_p$ is the surface charge of a platelet
per unit area.
Expressions of the free energy can be derived from eq. (\ref{eq:deltaF}) by
considering various real or virtual charging processes.
In practice, we have calculated $F$ using a constant Debye length 
charging process, from a situation where the platelet is neutral
$(Z'=0)$ to its final charge $(Z'=Z)$. For uncharged platelets,
$N^+_{_{0}} = N^-_{_{0}}=N_{_{0}} = \Omega \,n_s +Z/2$, where $n_s$ is the
salt concentration. At any stage of the process, $N^\pm = N_{_{0}} \pm 
Z'/2$, and eq. (\ref{eq:deltaF}) shows that in an infinitesimal
step, during which the boundary condition of vanishing 
normal electric field is enforced, the free energy changes by:
\begin{equation}
\beta\delta\left( F\right) = \int_{\Sigma_p} \Phi \, \delta \left(\frac{\sigma}{e}\right) \,  d^2{\bf r}
\,+\, \left[\ln\left(\rho_{_{0}}^+\Lambda_+^3\right) \right]\delta N_+ \,+\,
 \left[\ln\left(\rho_{_{0}}^-\Lambda_-^3\right) \right]\delta N_-.
\label{eq:deltaFsigma}
\end{equation}
Integration along this path yields:
\begin{eqnarray}
F(\sigma) - F\left(\sigma=0\right) &&=  
 \int_{0}^{\sigma} \left[ \int_{\Sigma_p} 
\varphi^{\sigma'}({\bf r}) \, d^2 {\bf r} \right] d\sigma' +
\nonumber \\
 &&N_{_{0}} kT\,\ln\left\{
\frac{(N_{_{0}})^2 - Z^2/4}{(N_{_{0}})^2}\right\} + 
\frac{Z}{2} kT\,\ln \left\{\frac{N_{_{0}} +Z/2}{N_{_{0}} -Z/2}\right\},
\label{eq:energie}
\end{eqnarray}
The result (\ref{eq:deltaF}), valid for a given WS cell geometry, may
be generalized to the case of an infinitesimal change $\delta \Omega$
of the geometry (shape and/or volume) of the cell. A calculation 
given in the appendix shows that for fixed numbers of co- and 
counterions ($N^+$ and $N^-$) in the cell, the resulting infinitesimal 
change in free energy reads:
\begin{equation}
\delta F = -k_{_{B}} T \sum_{\alpha = +,-} \int_{\delta \Omega} \,
\rho^{\alpha}({\bf r}) \, d^3{\bf r} \,- \, \frac{\varepsilon}{8\pi} 
\int_{\delta \Omega} \, \left[\bbox{\nabla} \varphi\right]^2 \, d^3{\bf r},
\label{eq:dFshape}
\end{equation}
where it has been assumed that the electric field has no normal component
on the WS surface. Under these circumstances, the previous expression
for $\delta F$ is compatible with the usual definition of the pressure
tensor in a charged medium \cite{LaLi}
\begin{equation}
\stackrel{\longleftrightarrow}{\bbox{\Pi}}=
\left[P({\bf r}) +\frac{\varepsilon}{8\pi}\left(
\bbox{E}\right)^2\right] 
\stackrel{\leftrightarrow}{\bbox{I}}
\,- \, \frac{\varepsilon}{4\pi} \,\bbox{E}({\bf r})
\otimes
\bbox{E}({\bf r}),
\label{eq:ptenseur}
\end{equation}
where $\bbox{E}({\bf r}) = -\bbox{\nabla}\varphi({\bf r})$, and for 
non-interacting ions,
\begin{equation}
P({\bf r}) = k_{_{B}}T \sum_{\alpha = +,-}\rho^{\alpha}({\bf r}).
\label{eq:pideal}
\end{equation}
The compatibility of eqs. (\ref{eq:dFshape}) and (\ref{eq:ptenseur})
plus (\ref{eq:pideal}) is easily verified by considering 
a small local volume change $\delta \Omega$ due to the displacement
by an amount $\delta \bbox{\ell}$ of a small area element $\delta
\bbox{\Sigma}$ at the surface of the WS cell. The pressure 
tensor (\ref{eq:ptenseur}) satisfies the mechanical equilibrium condition
\begin{equation}
\bbox{\nabla}\cdot \stackrel{\longleftrightarrow}{\bbox{\Pi}}
({\bf r}) = \bbox{0}
\label{eq:equil}
\end{equation}
which, upon substitution of (\ref{eq:ptenseur}) into 
(\ref{eq:equil}), and use of Poisson's equation (\ref{eq:poisso}),
is equivalent to the familiar force balance equation 
\begin{equation}
\bbox{\nabla} P({\bf r}) = e\,\rho_c({\bf r}) \, \bbox{E}({\bf r}).
\end{equation}

There is no obvious definition of the macroscopic
osmotic pressure of the co- and
counterions within the present WS model. An element 
$\bbox{d S}$ of the surface of the WS cell is subjected to the force:
\begin{equation}
\bbox{d F} = \stackrel{\longleftrightarrow}{\bbox{\Pi}}\!
\cdot  \, \bbox{dS}
= \left(k_{_{B}}T \sum_{\alpha = +,-} \rho^\alpha 
+\frac{\varepsilon}{8\pi} E^2\right)\bbox{d S}.
\end{equation}
It is hence tempting to define the osmotic pressure as:
\begin{equation}
\Pi = k_{_{B}}T \sum_{\alpha = +,-} \overline{\rho^\alpha}^{\,\Sigma}
\,+\,\frac{\varepsilon}{8\pi} \,\overline{E^2}^{\,\Sigma}
\label{eq:pression}
\end{equation}
where the averages are taken over the total surface bounding the WS
cell. The same expression for $\Pi$ follows from the volume derivative
of the free energy, for a particular infinitesimal dilation of the WS
cell. The latter is chosen such that for each surface element $dS$ of the
WS cell, the infinitesimal volume element is $d \Omega$ = $d \Sigma \, 
\delta \ell$, where $\delta \ell$ is a constant displacement along the 
normal to $d \Sigma$. Under these conditions, we deduce from eq.
(\ref{eq:dFshape}) that:
\begin{equation}
\delta F = -k_{_{B}}T\, \delta \ell \sum_{\alpha = +,-} 
\oint_{\Sigma} \rho^\alpha({\bf r}) \,d\Sigma - \frac{\varepsilon}{8\pi}\,
\delta \ell \oint_{\Sigma} \left(\bbox{\nabla} \varphi\right)^2 \, d\Sigma,
\end{equation}
so that:
\begin{equation}
\frac{\partial F}{\partial \Omega} = \lim_{\delta \ell \to 0} \frac{\delta F}
{\Sigma \,\delta \ell} = -k_{_{B}}T\, \overline{\rho^\alpha}^{\,\Sigma}
- \frac{\varepsilon}{8\pi}\, 
\overline{\left[\left(\bbox{\nabla} \varphi\right)^2\right]}^\Sigma = -\Pi.
\end{equation}
Similarly, one may express the disjoining pressure $\Pi_d$ ({\it i. e.}
the pressure to be applied to maintain the parallel platelets 
of a stack at a distance $H$ which coincides with the height of the WS
cell) by considering the variation of the free energy (\ref{eq:dFshape})
upon increasing $H$ by an infinitesimal amount $\delta H$; the 
corresponding increase in the volume of the WS cell is $\delta \Omega
= S\delta H$, where $S$ is the cross-section of the cell parallel to the 
platelet (cf. the prismatic geometries
considered in the following sections). By proceding as in the case 
of the osmotic pressure, one arrives at the required expression
\begin{equation}
\Pi_d = - 
\frac{\partial (F/S)}{\partial H} = 
k_{_{B}}T\, \overline{\rho^\alpha}^{\,S}
\,+ \,\frac{\varepsilon}{8\pi}\, 
\overline{\left[\left(\bbox{\nabla} \varphi\right)^2\right]}^S = 
\overline{\Pi_{zz}}^S,
\label{eq:disj}
\end{equation}
where the average is now taken over the cross-section $S$, 
{\it i. e.} on the surface 
of the WS cell parallel to the platelet,
and the normal to the platelet is 
chosen along the $z$ axis. $\Pi_d$ is thus related to the mean uniaxal
stress along the normal to the platelets.

Another instructive quantity is the capacitance $C$ of the electric
double-layer associated with a platelet. For finite platelets, the 
definition of $C$ is ambiguous. We define 
the capacitance in terms 
of the difference 
$\left( \Delta \overline{\varphi}\right)^{S'-S}=
\overline{\varphi}^{S'}
-\overline{\varphi}^S$ between the potential
averaged over the cross-section $S'$ of the cell containing the 
platelet ($z=0)$, and the corresponding 
average on the surface $S$ midway between two platelets in a stack 
($z=H/2$, 
cf. Fig. 1 for the case
of cylindrical and parallelepipedic WS cells).
The capacitance is thus defined by 
\begin{equation}
C\, \left( \Delta \overline{\varphi}\right)^{S'-S} = \sigma.
\label{eq:capa}
\end{equation}
The reciprocal of the capacitance defines a length $\lambda_c = C^{-1}$
which characterizes the thickness of the electric double-layer.

The total charge inside a WS cell is zero, and due to space reflection
symmetry, the electric dipole moment associated with the charge distribution 
inside the cell vanishes. The first {\it a priori} non vanishing 
multipole moment of the charge distribution is the quadrupole
moment $Q=Q_{zz}=-2Q_{xx}=-2Q_{yy}$:
\begin{equation}
Q_{zz}^{\hbox{\scriptsize tot.}} = \frac{1}{2}\int_{\Omega} 
\left\{q_{_{\cal P}}({\bf r}) + e\left[\rho^+({\bf r}) - 
\rho^-({\bf r})\right]
\right\}
\left(2z^2-x^2-y^2\right)\, d^3 {\bf r}.
\label{eq:quadru}
\end{equation}
In the two following sections, the quantities defined in this section
will be calculated within cylindrical and parallelepipedic geometries,
for disc-shaped and square platelets.

\section{Cylindrical geometry}
We consider first the case of disc-shaped platelets, of radius 
$r_{\scriptscriptstyle 0}$.
In ref. \cite{HaTr} explicit solutions of LPB theory were obtained
in terms of infinite series of Legendre or Bessel functions, for spherical
and cylindrical WS cells. We reexamine the latter geometry in some detail
here. The cylindrical WS cell is of radius $R$ and of height $H=2h$ 
(cf. Fig. 1), so that $\Omega=2\pi \, R^2\, h$. The electrostatic potential
within the WS cell is a function $\varphi(r,z)$ of the cylindrical 
coordinates $r$ and $z$, which satisfies the boundary conditions
\begin{mathletters}
\begin{eqnarray}
\label{eq:bca}
& & \frac{\partial \varphi(r,z)}{\partial r} \biggl|_{r=R} = 0, \\
& & \frac{\partial \varphi(r,z)}{\partial z} \biggl|_{z=\pm h} = 0.
\label{eq:bcb}
\end{eqnarray}
\end{mathletters}
Since these conditions, as well as the discontinuity of the electric field
$E_{z}(z=0^\pm) = -(\partial \varphi/\partial z)_{z=0^\pm}$ across
the disc ($r<r_{\scriptscriptstyle 0}$), involve only the derivatives
of the potential, one may assume $\overline{\varphi}=0$ without loss
of generality (cf. eq. (\ref{eq:phibarre})). Under these conditions, 
the solution of the linearized  PB equation (\ref{eq:dh})
may be expanded in a Bessel-Dini series \cite{HaTr}:
\begin{equation}
\varphi(r,z)= \sum_{n=1}^{\infty} \, A_{n}(z) \, J_{0}\left(y_{n} \frac{r}{R}
\right),
\label{eq:dini}
\end{equation}
where $y_n$ is the $n^{th}$ root of $J_{1}(y) \equiv -dJ_{0}(y)/dy=0$, and
$J_{0}$ and $J_{1}$ are the Bessel functions of $0^{th}$ and $1^{st}$
order ($y_1=0$).
The resulting differential equations for the coefficients $A_n(z)$ are easily
solved, leading to the final result:
\begin{eqnarray}
\Phi(r,z) = \beta e \, \varphi(r,z) 
     &=&  \beta e \gamma_{\scriptscriptstyle 0} \, + 
\, \frac{1}{\kappa_{_{D}} b} \, 
\left(\frac{r_{\scriptscriptstyle 0}}
{R}\right)^{2} \, \, \frac{\cosh[\kappa_{_{D}}(h-|z|)]}{\sinh(\kappa_{_{D}}h)}
\, \nonumber \\
 & & \mbox{} + \frac{2}{b} \frac{r_{\scriptscriptstyle 0}}{R} \sum_{n=2}^{\infty} 
\frac{\Lambda_{n} \, J_{1}(k_{n} r_{\scriptscriptstyle 0}) }{ y_{n} \, \sinh\left(h/
\Lambda_{n}\right) \, J_{0}^{2}(y_{n}) } 
\cosh \left(\frac{h-|z|}{\Lambda_{n}}
\right) \, J_{0}(k_{n} r), 
\label{eq:seriecyl}
\end{eqnarray}
where $b = e/(2\pi \ell_{\scriptscriptstyle B}\sigma )$ is the Gouy length,
$\Lambda_{n} = R/\sqrt{y_{n}^{2}+\kappa_{_{D}}^{2} R^{2}}$ and 
$k_{n}=y_{n}/R$.

For any given macroscopic density of platelets, $n$, and hence a given volume 
$\Omega = 1/n$ of the WS cell, only the product $R^2 h$ is fixed.
Since $R$ cannot be less than the disc radius
$r_{\scriptscriptstyle 0}$,
the aspect ratio $h/r_{\scriptscriptstyle 0} \leq \left(2\pi n 
r_{\scriptscriptstyle 0}^3\right)^{-1}$. At the upper limit of this range,
{\it i. e.}  for 
$R=
r_{\scriptscriptstyle 0}
$, and with the chosen boundary conditions,
the electrostatic problem reduces to that of an infinite uniformly
charged plane in a WS slab of width $H=2h$. 
With $R=r_{\scriptscriptstyle 0}$
the potential in eq. (\ref{eq:seriecyl}) indeed reduces to the first 
two terms, which are independent of the radial coordinate $r$, {\it i. e.}:
\begin{equation}
\lim_{R\to r_{\scriptscriptstyle 0}} \Phi(r,z) = \Phi(z) =
\beta e \, \gamma_{\scriptscriptstyle 0} + \frac{1}{\kappa_{_{D}} b} \, 
\frac{\cosh[\kappa_{_{D}}(h-|z|)]}{\sinh(\kappa_{_{D}}h)}.
\label{eq:planinfini}
\end{equation}
The above expression is precisely the solution of the one-dimensional 
LPB equation for an
infinite uniformly charged plane in a slab, and the familiar
result $\Phi(z) = \exp(-\kappa_{_{D}} z)/(\kappa_{_{D}} b)$
is recovered when $h \to \infty$.

The density profiles calculated from the potential (\ref{eq:seriecyl})
via eq. (\ref{eq:phista}) are sensitive to the aspect ratio 
$h/r_{\scriptscriptstyle 0}$ for a given cell volume $\Omega$. The 
``optimum'' ratio is determined by minimizing the free energy $F$ 
with respect to this ratio. $F$ is calculated via the constant 
Debye length charging process, resulting in eq. (\ref{eq:energie}).
The part of that expression which depends on the ratio 
$h/r_{\scriptscriptstyle 0}$ for a given cell volume is:
\begin{eqnarray}
A = \int_{0}^{\sigma} \left[ \int_{\Sigma_p} 
\varphi^{\sigma'}(r,z=0) \, d^2 {\bf r} \right] d\sigma' &=&
2\pi \int_{0}^{\sigma} \left[ \int_{0}^{r_{\scriptscriptstyle 0}}
\varphi_{\scriptscriptstyle 0}^{\sigma'}(r) \, r\, d r \right] d\sigma'
\nonumber \\
&=& \pi \sigma \int_{0}^{r_{\scriptscriptstyle 0}}
\varphi_{\scriptscriptstyle 0}^{\sigma}(r) \, r\, d r 
\label{eq:acyl}
\end{eqnarray}
where $\varphi_{\scriptscriptstyle 0}(r) \equiv \varphi(r,z=0)$ is the
electrostatic potential on the disc and the last line holds within
the LPB approximation only, where the potential
can always be expressed in the form
$\varphi^\sigma({\bf r})
= \sigma \, f(\kappa_{_{D}},{\bf r})$.
Substitution of eq. (\ref{eq:seriecyl}) (with $z=0$) into eq.
(\ref{eq:acyl}) yields
\begin{equation}
\frac{\beta}{Z}\, A = \frac{1}{2} \,\beta e \gamma_{\scriptscriptstyle 0}-
\frac{1}{2 \kappa_{_{D}} b} \left\{\left(\frac{r_{\scriptscriptstyle 0}}{R}\right)^2\frac{1}
{\hbox{tanh}(\kappa_{_{D}} h)} + 4\sum_{n=2}^{\infty} 
\frac{\kappa_{_{D}} \Lambda_n \, J_1^2\left(k_n r_{\scriptscriptstyle 0}\right)}
{y_n^2\,J_0^2(y_n)\,\hbox{tanh}(h/\Lambda_n)}\right\}.
\label{eq:fdisccyl}
\end{equation}
An example of the variation of $F$ with $h/r_{\scriptscriptstyle 0}$
is shown in Fig. 2a. In this example, as well as under all physical
conditions that were investigated, $F$ goes through a minimum
for a ratio $h/r_{\scriptscriptstyle 0}$, such that the physical 
requirement that $R>r_{\scriptscriptstyle 0}$ is satisfied. Moreover,
the location of the minimum turns out to be practically 
independent of the charge density $\sigma$ carried by the disc,
and of the salt concentration $n_s$; 
in other words, the ratio
$h/r_{\scriptscriptstyle 0}$ depends ``only'' on the WS cell volume 
$\Omega$, or equivalently on the macroscopic density of platelets $n$.
For any density $n$, the system selects an optimal ratio
$h/r_{\scriptscriptstyle 0}$. The variation of this ratio with $n$ 
in shown in Fig. 3. All figures correspond to 
$\varepsilon_{\hbox{\scriptsize CGS}} = 78$ and $T=300\,K$.

The potential $\varphi(r,z)$ and the resulting charge density 
profile $\rho_c(r,z)$ are highly anisotropic functions, as expected
from the platelet geometry. This requires that a large number of terms
(typically 50 to 400) be retained in the expansion (\ref{eq:seriecyl}) 
to ensure adequate convergence. A typical example of equipotential lines
is shown in Fig. 4. 

If $n'_s$ is the salt concentration in a reservoir which is in osmotic
equilibrium with the colloidal suspension, {\it i. e.} with the co- and
counterions inside the WS cell, then the salt concentration $n_s$
inside the latter is, within LPB theory, related to $n'_s$ by \cite{HaTr}:
\begin{equation}
\frac{n_s}{n'_s} =  \sqrt{1+\frac{Z^2\,n^2}{4(n'_s)^2}} - 
\frac{Zn}{2n'_s} \,\leq \,1,
\end{equation}
which is an expression of the familiar Donnan effect.

The values of the potential and its gradient on the surface $\Sigma$ of
the WS cell may be used to compute the pressures $\Pi$ and 
$\Pi_d$, according to equations (\ref{eq:pression}) and (\ref{eq:disj}).
For a given WS cell volume $\Omega$, the variations of $\Pi$ and
$\Pi_d$ with the ratio $h/r_{\scriptscriptstyle 0}$ differ:
$\Pi_d$ decreases monotonously as $h/r_{\scriptscriptstyle 0}$ increases,
while $\Pi$ goes through a minimum for $h/r_{\scriptscriptstyle 0}$
close to the ``optimum'' value minimizing the free energy (cf Fig. 2).
At the minimum of the free energy, the two pressures are seen to coincide
($\Pi=\Pi_d$). Taking into account
the conservation of the overall volume 
$\delta \left(R^2 h\right) = 0$, eq. (\ref{eq:dFshape}) can indeed 
be rewritten as
\begin{equation}
\delta F = \Sigma \,\frac{R}{2} \, \frac{\delta h}{h} \left[\Pi-\Pi_d\right],
\label{eq:Fmin}
\end{equation}
so that the extremum condition $\delta F=0$ implies $\Pi=\Pi_d$.
The variation of $\Pi=\Pi_d$ with the platelet concentration $n$
and with the salt concentration is shown in Figures 5a and 5b for 
two values of the platelet charge ($Z=100$ and $Z=200$). 
After a shallow minimum, the pressure $\Pi$ is found to increase with $n$
(for fixed $n'_s$ in the reservoir), whereas it drops rapidly 
with increasing $n'_s$, towards the value of the osmotic pressure
in the reservoir. These tendencies are reminiscent of the experimental
and numerical results of Dubois and coworkers concerning 
lamellar phases of ``infinite'' charged bilayers \cite{DZBD}.

The potential distribution may be used to evaluate the capacity
$C$ from equation (\ref{eq:capa}). The characteristic double-layer
thickness is easily calculated to be:
\begin{equation}
\lambda_c = \lambda_{_{D}} \, \left(\frac{
r_{\scriptscriptstyle 0}
}{R}\right)^2 \, \tanh\left(\frac{h}{2\lambda_{_{D}}}\right).
\label{eq:lambda}
\end{equation}
The variation of $\lambda_c$ with platelet and salt concentrations
is illustrated in Fig. 6a and 6b, for two values of the platelet 
charge. $\lambda_c$ reduces to the Debye length $\lambda_{_{D}}$ 
when $r_{\scriptscriptstyle 0} \to R$ and $h\to \infty$, which
corresponds to the limit of a single uniformly charged infinite 
plane.

Finally, as illustrated in Fig. 2, the quadrupole moment
of the charge distribution inside the WS cell vanishes at the 
``optimum'' ratio $h/r_{\scriptscriptstyle 0}$, which minimizes the free 
energy. This coincidence is systematic and may be related to an exact
result of Gruber et al. \cite{GrLM}, provided the WS model
yields an accurate description of a regular, periodic stacking
of the platelets \cite{TrHa2}. A partial explanation follows from
the calculation given at the end of the appendix. When the
Boltzmann's weights (\ref{eq:boltzm}) are linearized, 
the kinetic part of the free energy variation (cf. eq. (\ref{eq:dFpartitiona}))
as a function of the aspect ratio is proportional
to the total quadrupole (cf. eq. (\ref{eq:Fcinquadru})). 
Fig. 2b shows this correspondance but also shows that 
the electrostatic part of the free energy variation, defined from
eq. (\ref{eq:dFpartitionb}), vanishes for the same 
aspect ratio as the kinetic one.

\section{Parallelepipedic geometry}

Instead of a cylindrical WS cell, we now consider a space-filling
parallelepipedic cell of dimensions $L\!*\!L\!*\!H$
in the $x$, $y$ and
$z$ directions respectively (cf Fig. 1). 
The potential is naturally expanded in plane
waves, compatible with the periodic boundary conditions, which
are equivalent to the condition of vanishing normal component
of the electric field on the surface $\Sigma$ bounding 
the prismatic WS cell:
\begin{equation}
\varphi({\bf r}) = \sum_{{\bf k}} \widetilde{\varphi}\left({\bf k}\right)\,
\exp\{i\, {\bf k}\!\cdot \!{\bf r}\},
\end{equation}
with
\begin{equation}
{\bf k} = 2\pi \left(\frac{n_x}{L},\frac{n_y}{L},\frac{n_z}{H}\right), 
\qquad (n_x, n_y, n_z) \in \hbox{\it Z\hskip -4pt Z}\,^3.
\end{equation}
In terms of Fourier components, eq. (\ref{eq:dh}) becomes
\begin{equation}
\left(k^2+\kappa_{_{D}}^2\right) \left(\widetilde{\varphi}({\bf k}) - \gamma_{\scriptscriptstyle 0} \,\delta_{\bbox{k},\bbox{0}}
\right) = \frac{4\pi}{\varepsilon} \widetilde{q}_{_{\cal P}}({\bf k}),
\label{eq:dhfourier}
\end{equation}
where
\begin{equation}
\widetilde{q}_{_{\cal P}}({\bf k}) = \frac{1}{\Omega} \int_{\Omega} 
q_{_{\cal P}}({\bf r}) \, \exp\{i\, {\bf k}\!\cdot\!{\bf r}\} \, d^3 {\bf r}.
\end{equation}

Consider first the case of a circular platelet (or disc) of radius
$r_{\scriptscriptstyle 0}$
, as in section IV; $\widetilde{q}_{_{\cal P}}({\bf k})$
is easily calculated to be
\begin{equation}
\widetilde{q}_{_{\cal P}}({\bf k}) = \frac{2\pi r_{\scriptscriptstyle 0}}{k_{\scriptscriptstyle /\!/}} \, 
J_1\left(k_{\scriptscriptstyle /\!/} r_{\scriptscriptstyle 0}\right) 
\label{eq:rhotildis}
\end{equation}
with 
$$
k_{\scriptscriptstyle /\!/} = \frac{2\pi}{L} \sqrt{n_x^2+n_y^2}.
$$
The Fourier components of the electrostatic potential are then
determined by substituting eq. (\ref{eq:rhotildis}) into eq. 
(\ref{eq:dhfourier}), and inverse Fourier transformation 
leads to the desired
result:
\begin{eqnarray}
\Phi({\bf r}) = 
\beta e\, \varphi({\bf r}) = \beta e  \gamma_{\scriptscriptstyle 0}+ \frac{2\pi}{b}
\frac{r_{\scriptscriptstyle 0}}{L^2}\,
 \sum_{(n_x, n_y) \, \in\, \hbox{\scriptsize \it Z\hskip -4pt Z}\,^2\,}
\frac{J_1\left(k_{\scriptscriptstyle /\!/} r_{\scriptscriptstyle 0}\right) 
}{k_{\scriptscriptstyle /\!/}} \, \cos\left(k_x \,x+k_y\, y\right) 
\nonumber\\
\frac{1}{(\kappa_{_{D}}^2+
k_{\scriptscriptstyle /\!/}^2)^{1/2}}\,
\frac{\cosh\left[(\kappa_{_{D}}^2+k_{\scriptscriptstyle /\!/}^2)^{1/2}
(h-z)\right]}
{\sinh\left[h(\kappa_{_{D}}^2+k_{\scriptscriptstyle /\!/}^2)^{1/2}\right]}.
\label{eq:potdispar}
\end{eqnarray}
The resulting free energy, as calculated from the constant 
Debye length charging process (\ref{eq:energie}) and
(\ref{eq:acyl}), is:
\begin{equation}
\frac{\beta}{Z}\, A = \frac{1}{2} \,\beta e \gamma_{\scriptscriptstyle 0}
-\frac{4\pi}{L^2H b}\,
\sum_{(n_x, n_y, n_z) \, \in\, \hbox{\scriptsize \it Z\hskip -4pt Z}\,^3\,}
\frac{J_1^2\left(k_{\scriptscriptstyle /\!/} r_{\scriptscriptstyle 0}\right)}{k_{\scriptscriptstyle /\!/}^2\left(k^2+\kappa_{_{D}}^2\right)}.
\label{eq:adiscpar}
\end{equation}
Similarly, the quadrupolar moment may be calculated from eq. 
(\ref{eq:potdispar}) to be
\begin{equation}
\frac{Q_{zz}^{\hbox{\scriptsize tot.}}}{Q_{\hbox{\scriptsize disc}}} =
-\frac{16}{r_{\scriptscriptstyle 0}^3}\left\{\sum_{n_z=1}^{\infty} \frac{(-1)^{n_z}}
{k_z^2+\kappa_{_{D}}^2} \,r_{\scriptscriptstyle 0} - 
\sum_{n_x=0}^{\infty} \frac{J_1\left(k_x r_{\scriptscriptstyle 0}\right)}{k_x^2+\kappa_{_{D}}^2}(-1)^{n_x}
\frac{2}{k_x}\right\}.
\label{eq:qdiscpar}
\end{equation}

The key finding is that, for given platelet and salt concentrations, 
the changes in free energy and quadrupole moment, induced by the change of topology of the WS cell, are practically negligible, in spite of the
completely different analytical expressions (cf. eq.
(\ref{eq:acyl}) and (\ref{eq:adiscpar})), as illustrated in Fig. 7.
A similar conclusion holds for the osmotic pressure calculated
from eq. (\ref{eq:pression})
or the capacity evaluated from eq. (\ref{eq:capa}).

We finally consider the case of a square platelet of side 
$l_{\scriptscriptstyle 0}$, within the above prismatic WS cell. 
In this case, the surface charge per unit area is 
$\sigma = -Ze/l_{\scriptscriptstyle 0}^2$, while the Fourier transform of the 
platelet charge density is now
\begin{equation}
\widetilde{q}_{_{\cal P}}({\bf k}) = 
\frac{\sigma}{\Omega} \,\frac{4}{k_x k_y}\,\sin\left(\frac{k_x
l_{\scriptscriptstyle 0}}{2}\right) \,\sin\left(\frac{k_y
l_{\scriptscriptstyle 0}}{2}\right).
\end{equation}
The potential is once more expanded in plane waves with the result
\begin{eqnarray}
\Phi({\bf r}) = 
\beta e \,\varphi({\bf r}) = \beta e \gamma_{\scriptscriptstyle 0} +
\frac{4}{b L^2}\, \sum_{(n_x, n_y) \, \in\, 
\hbox{\scriptsize \it Z\hskip -4pt Z}\,^2\,}
\sin\left(k_xl_{\scriptscriptstyle 0}/2\right) \sin\left(k_yl_{\scriptscriptstyle 0}/2\right)
\frac{1}{k_x k_y} 
\nonumber\\
\cos\left(k_x \,x+k_y\, y\right) \,\frac{1}{(\kappa_{_{D}}^2+
k_{\scriptscriptstyle /\!/}^2)^{1/2}}\,
\frac{\cosh\left[(\kappa_{_{D}}^2+k_{\scriptscriptstyle /\!/}^2)^{1/2}
(h-z)\right]}
{\sinh\left[h(\kappa_{_{D}}^2+k_{\scriptscriptstyle /\!/}^2)^{1/2}\right]}.
\label{eq:potcar}
\end{eqnarray}
As in the case of a disc in a cylindrical 
WS cell, one can check from eq. (\ref{eq:potcar}) that the potential
goes over to that of a uniformly charged infinite plane
when $l_{\scriptscriptstyle 0} \to L$ (expression (\ref{eq:planinfini})).

The corresponding free energy is now:
\begin{equation}
\frac{\beta}{Z}\, A = \frac{1}{2} \,\beta e \gamma_{\scriptscriptstyle 0}
-\frac{16}{l_{\scriptscriptstyle 0}^2 L^2 H b}\,
\sum_{(n_x, n_y, n_z) \, \in\, \hbox{\scriptsize \it Z\hskip -4pt Z}\,^3\,} 
\frac{\sin^2\left(k_x l_{\scriptscriptstyle 0}/2\right)\, 
\sin^2\left(k_y l_{\scriptscriptstyle 0}/2\right)}{k_x^2\, k_y^2\left(k^2+\kappa_{_{D}}^2\right)},
\end{equation}
while the quadrupole moment reads
\begin{equation}
\frac{Q_{zz}^{\hbox{\scriptsize tot.}}}{Q_{\hbox{\scriptsize platelet}}} =
-\frac{48}{l_{\scriptscriptstyle 0}^3} \left\{\sum_{n_z=1}^{\infty} (-1)^{n_z} 
\frac{l_{\scriptscriptstyle 0}}{k_z^2+\kappa_{_{D}}^2} - 2 \sum_{n_x=1}^{\infty} (-1)^{n_x} 
\frac{\sin(k_x l_{\scriptscriptstyle 0}/2)}{k_x\left(k_x^2+\kappa_{_{D}}^2\right)}\right\}.
\end{equation}
Explicit calculations based on these formulae lead again to results
which are quite close to those obtained for circular platelets
under similar physical conditions, as shown {\it e. g.} 
in Fig. 7. In particular, the ``optimum'' aspect ratio  is characterized
by the equality of the osmotic and disjoining pressures (an expression similar
to eq. (\ref{eq:Fmin}) holds), as illustrated 
in Fig. 8. However, the minima of the free energy and
of the osmotic pressure do not coincide any more.

\section{Infinite dilution limit}

The limit of very low platelet concentration $n$, for a fixed 
salt concentration $n_s$ (or equivalently $n'_s$) may be derived
from the expressions obtained in sections IV and V for 
prismatic geometries, by letting the volume $\Omega$
of the WS cell go to infinity. In this limit, 
$\gamma_{\scriptscriptstyle 0}=0$, and
the series (\ref{eq:seriecyl}) and (\ref{eq:potdispar}) 
go over to the following 
integral representation of the reduced potential
\begin{equation}
\Phi(r,z) = \frac{r_{0}}{b} \, \int_{0}^{\infty} \, dk \,
J_{0}(k r)\, J_{1}(k r_{0}) \, \frac{ e^{-\sqrt{\kappa_{_{D}}^{2}+k^{2}}|z|}}
{\sqrt{\kappa_{_{D}}^{2}+k^{2}} }.
\label{eq:infini}
\end{equation}
Along the $z$-axis passing through the centre of the disc
($r=0$), eq. (\ref{eq:infini}) reduces to:
\begin{equation}
\Phi(r=0,z) =
\frac{1}{\kappa_{_{D}} b} \, \left[ \, e^{-\kappa_{_{D}} |z|}
-e^{-\kappa_{_{D}} \sqrt{r_{0}^{2}+z^{2}}} \,\right],
\label{eq:plan}
\end{equation}
which goes over to the familiar exponential solution 
of linearized Gouy-Chapman theory in the limit 
$r_{\scriptscriptstyle 0} \to \infty$ (infinite plane). 
The reduced potential at the centre of the disc takes the value
\begin{equation}
\Phi(\bbox{0})= \frac{e \varphi(r=0,z=0)}{k_{_{B}}T} = 
\frac{1}{\kappa_{_{D}} b} \left\{1-\exp(-\kappa_{_{D}}r_{0})\right\}.
\label{eq:potinf0}
\end{equation}
This expression will be used in the concluding section to evaluate the range
of validity of LPB theory.

The quadrupole moment around circular or square platelets vanishes
identically in the zero concentration limit $(n \to 0)$ as a  
consequence of the theorem by Gruber {\it et al.} \cite{GrLM}.

Finally consider the problem of determining the force acting
between two circular platelets (discs) ${\cal P}_1$ and ${\cal P}_2$,
in the limit of vanishing concentration ($n\to 0$), and for
a given salt concentration (in the limit $n\to 0$, 
$n_s \to n'_s$). Let $({\bf r}_1, {\bf n}_1)$
and $({\bf r}_2, {\bf n}_2)$ be the centre positions and normals
of the two discs, and let $\varphi({\bf r})$
be the total electrostatic potential due to the two discs and
their electric double-layers; $\varphi({\bf r})$ vanishes as 
$|{\bf r}| \to \infty$. In the LPB approximation,
$\varphi({\bf r})$ satisfies eq (\ref{eq:dh}), with 
$\gamma_{\scriptscriptstyle 0}=0$, and two source terms
\begin{equation}
\left(\nabla^{2}  -
\kappa_{_{D}}^{2}\right) \varphi({\bf r}) =
-\frac{4 \pi}{\varepsilon}\,\left[ q_{_{{\cal P}_1}}({\bf r}) +
q_{_{{\cal P}_2}}({\bf r}) \right].
\label{eq:dh2}
\end{equation}
Since the discontinuity of the electric field upon crossing one of 
the platelets (characterized by the surface charge density 
$ q_{_{\cal P_\alpha}}({\bf r})$) is independent of the presence
of the other platelet, the solution $\varphi({\bf r})$ 
of eq. (\ref{eq:dh2}) is just the superposition of the solutions
for each platelet separately:
\begin{equation}
\varphi({\bf r}) = \varphi_1({\bf r}) + \varphi_2({\bf r}).
\end{equation}
This simple property is a consequence of the linearization, and does
not hold within non-linear PB theory. The solution of eq. (\ref{eq:dh2}),
and the resulting co- and counterion density profiles may now be used to 
calculate the pressure tensor (\ref{eq:ptenseur}) for given positions
and orientations of the platelets. The force ${\bf F}_i$
acting on platelet $i$ (=1 or 2) follows then from integration of 
$\stackrel{\longleftrightarrow}{\bbox{\Pi}}$ over the two faces
$\Sigma_p^+$ and $\Sigma_p^-$ of the platelet:
\begin{equation}
{\bf F}_i = -\int_{\Sigma_{p,i}^+ ;\, \Sigma_{p,i}^-} 
\stackrel{\longleftrightarrow}{\bbox{\Pi}}\!\cdot\, {\bf dS}_i.
\label{eq:force}
\end{equation}
Since the normal ${\bf n}_i = {\bf d S}_i/|{\bf d S}_i|$ has opposite 
orientations on the faces 
$\Sigma_{p,i}^+$ and $\Sigma_{p,i}^-$, 
the kinetic contributions $\rho^\alpha({\bf r})\, k_{_{B}}T$ to the pressure
tensor (\ref{eq:ptenseur}), which are continuous across the platelet,
do not contribute to the force (\ref{eq:force}). The normal
component of the electric field ${\bf E} = -\bbox{\nabla} \varphi$, 
however, suffers a discontinuity across the uniformly charged platelet, 
and hence
contributes to the surface integral in eq. (\ref{eq:force}). 

The total electric field ${\bf E}$ in the immediate vicinity of the platelet
${\cal P}_i$ may be decomposed into a discontinuous and a continuous
part:
\begin{equation}
{\bf E} = {\bf E}_i^{(d)}+{\bf E}_i^{(c)} = 
\pm \frac{2\pi}{\varepsilon} \, \sigma \,
{\bf n}_i + {\bf E}_i^{(c)}
\label{eq:decomp}
\end{equation}
where the $+$ and $-$ signs go with the upper ($\Sigma_{p,i}^+$)
and lower ($\Sigma_{p,i}^-$) faces respectively. 
Subsituting eq. (\ref{eq:decomp}) into (\ref{eq:force})
leads to the desired result:
\begin{equation}
{\bf F}_i = \sigma \int_{\Sigma_{p,i}} {\bf E}_i^{(c)} \,d^2 S
\quad (i=1,2),
\label{eq:forcefin}
\end{equation}
where the integration is now over the surface of the platelet.
For symmetry reasons, the only non-vanishing contribution 
of ${\bf E}_i^{(c)}$ to the surface integral in eq. (\ref{eq:forcefin})
is the electric field due to the other platelet and its associated
electric double-layer. In the case of two coaxial parallel discs,
the force ${\bf F}_1 = -{\bf F}_2$
is along the common axis (chosen to be the $z$-axis) and the result
(\ref{eq:infini}) may be used to compute the gradient of $\varphi$ along
$Oz$. The resulting force is easily cast in the form:
\begin{equation}
F_z(d) = (\pi r_{\scriptscriptstyle 0}^2) \frac{4\pi \sigma^2}{\varepsilon}
\int_{0}^{\infty} J_1^2(x) \,\frac{1}{x} \,\exp\left\{
-\frac{d}{r_{\scriptscriptstyle 0}}
\sqrt{
x^2+\kappa_{_{D}}^2 r_{\scriptscriptstyle 0}^2}
\right\}\, dx,
\label{eq:forcez}
\end{equation}
where $d$ is the distance between the two discs.
For any finite salt concentration ({\it i. e.} non-vanishing
$\kappa_{_{D}}$), the decay of $F_z$ with $d\,$ is essentially 
exponential as illustrated in Fig. 9. In the limit of vanishing
salt concentration $(\kappa_{_{D}} \to 0)$, $F_z$ decays like
a power-law. For $d \gg r_{\scriptscriptstyle 0}$, we find in that
limit
\begin{equation}
F_z(d) ~\stackrel{d \gg r_{\scriptscriptstyle 0}}{\sim}~ 
\frac{1}{\varepsilon}
\left(\frac{\pi r_{\scriptscriptstyle 0}^2 \sigma}{d}\right)^2
\left[1-\frac{3 \,r_{\scriptscriptstyle 0}^2}{2 \,d^2}\right].
\end{equation}
Note that the force (\ref{eq:forcez}) is repulsive at all distances.

\section{Conclusion}
While most of the existing theoretical literature on suspensions
of charged lamellar particles or membranes deals with the simpler 
one-dimensional problem of infinite charged planes, we have examined 
in this paper the case of stacks of finite-size, circular
or square platelets corresponding, for instance, to swollen clays.
The intractable many-platelet problem is reduced to the much simpler
problem of a single platelet within an electrically neutral
Wigner-Seitz cell of appropriate volume and topology. 
The co- and counterion density profiles 
have been obtained from analytic solutions of linearized 
Poisson-Boltzmann equation with appropriate boundary conditions
on the surface of cylindrical and parallelepipedic cells. The relevant
characteristics of the electric double-layer and the resulting 
thermodynamic properties have been calculated over a wide
range of physical conditions. The most notable results may be 
summarized as follows:

{\bf a)} For a given cell volume $\Omega$, the system selects an 
optimal size ratio corresponding to the minimum free energy.
The selected size ratio is practically independent of the platelet
charge density $\sigma$, and of the salt concentration, but varies
with the macroscopic platelet concentration $n$.

{\bf b)} The osmotic presure $\Pi$ and depletion pressure $\Pi_d$
due to the co- and counterions have qualitatively different variations
with the aspect ratio for fixed $\sigma$, clay and salt concentration,
but coincide at the optimum aspect ratio which minimizes the free 
energy. We have a simple explanation for this coincidence but not for
the observation that the osmotic pressure exhibits a minimum at the 
same aspect ratio as the free energy, at least in the case of 
circular platelets.

{\bf c)} The total quadrupole moment of the charge distribution 
in the WS cell vanishes at the optimal aspect ratio; this observation
may be related to an exact property of neutral charge distributions
in thermodynamic equilibrium \cite{GrLM}.
A partial explanation of this correlation has been given at the end
of section IV.

The limitations of the present LPB theory must be underlined.
Returning to the expression (\ref{eq:potinf0}) of the reduced
electrostatic potential at the centre of an {\em isolated}\/
charged disc, it is clear that linearization of the Boltzmann
factors in the PB equation (\ref{eq:pb}) is justified only provided
$|\Phi(\bbox{0})| \ll 1$, {\it i. e.} when $|b| > \lambda_{_{D}}$
(which corresponds to the limit of low surface charge $\sigma$
or high salt concentration),
or $r_{\scriptscriptstyle 0} < \lambda_{_{D}}$ and 
$r_{\scriptscriptstyle 0} < |b|$. This latter condition
is rather academic for Laponite clay
discs (it yields $Z<10$). 
Fig. 10 summarizes the limits of validity of the linearized
theory for an isolated platelet. Strictly speaking, the above
criteria apply in the infinite dilution limit $n\to 0$, 
and we may expect that they become necessary but not sufficient
conditions for finite concentrations $n$ \cite{Ande},
so that the shaded area in Fig. 10 should decrease. 
Neglecting the clay contribution 
to the Debye length, the criterion $\lambda_{_{D}} <|b|$ yields
$Z<2.3\, 10^{-n/2}\, r_{\scriptscriptstyle 0}^2$,
for a $10^{-n}$ molar monovalent salt concentration, with 
$r_{\scriptscriptstyle 0}$ expressed in nanometers.
In practice, we carried out
calculations for $r_{\scriptscriptstyle 0}=12.5\,$nm (a typical
size of Laponite particles); this means that $Z$ must be chosen less
than about $400\, 10^{-n/2}$. In fact, many of our calculations
were carried out for $Z=100$, in which case, strictly speaking,
the salt concentration would have to be $10^{-1}\,$M or higher for LPB 
theory to be applicable. However, the LPB results should
remain, at least qualitatively, valid at lower salt concentrations.
The discussion shows that more realistic calculations,
for values of $Z$ typical of clays ({\it e. g.} $Z\simeq 10^3$ for 
Laponite), and lower salt concentrations (to ensure stability
against flocculation) will require solutions of the full (non-linear)
PB equation within the WS geometry. Work along these lines
is in progress.

{\bf Acknowledgments:} Useful conversations with Marjolein
Dijkstra, Paul Madden and Thierry Biben 
are gratefully acknowledged.

\newpage
\begin{center}
{\bf \large APPENDIX}
\end{center}

Consider the case of a charged platelet ${\cal P}$ confined 
in a WS cell with its neutralizing counterions and salt. A local change 
$\delta \Omega$ in volume of the cell 
(with or without conservation of the overall
volume) changes the potential and the microion densities, but not
the charge density of the platelet. The boundary conditions
of vanishing normal electric field on the surface $\Sigma$ of the cell
are enforced during this change. It will be shown that when the total 
numbers of ions $N^\alpha$ in the cell are held constant, the free
energy $F$ changes according to eq. (\ref{eq:dFshape}). From eq.
(\ref{eq:u}), the change in the internal energy can be written:
\begin{equation}
\delta U = \frac{1}{2}\int_{\Omega} \left(q\, \delta \varphi + 
\varphi \,\delta q \right) d^3{\bf r} + 
\frac{1}{2} \int_{\delta \Omega} q\,\varphi \, d^3{\bf r},
\end{equation}
where $q({\bf r}) = q_{_{{\cal P}}}({\bf r}) + e[\rho^+({\bf r}) 
-\rho^-({\bf r})]$ is the total charge density. It follows from
Poisson's equation (\ref{eq:poisso}) and integrations by parts
that
\begin{equation}
\int_{\Omega} q \,\delta \varphi\, d^3{\bf r} = 
\int_{\Omega} \varphi \, \delta q\, d^3{\bf r} + \frac{\varepsilon}{4\pi}
\oint_{\Sigma} \varphi \, \bbox{\nabla}\left(\delta \varphi\right)\cdot
\bbox{dS}.
\end{equation}
For any vector field $\bbox{A}({\bf r})$, the elementary variation of
the flux over the surface $\Sigma$ can be written:
\begin{equation}
\delta \left(\oint_{\Sigma} \bbox{A}\cdot \bbox{dS} \right)= 
\oint_{\Sigma} \delta \bbox{A}\cdot \bbox{dS} + \int_{\delta \Omega} 
\bbox{\nabla}\!\cdot\!\bbox{A}\, d^3{\bf r}.
\end{equation}
With $\bbox{A} = \varphi \,\bbox{\nabla}\varphi$, the above relation 
can be used to 
differentiate the identity
\begin{equation}
\oint_{\Sigma}  \varphi \,\bbox{\nabla}\left(\varphi\right)\cdot
\bbox{dS} =0.
\end{equation}
One finds
\begin{equation}
\frac{\varepsilon}{4\pi}
\oint_{\Sigma} \varphi \, \bbox{\nabla}\left(\delta \varphi\right)\cdot
\bbox{dS} = \int_{\delta \Omega} q \, \varphi \, d^3{\bf r} -
\frac{\varepsilon}{4\pi} \int_{\delta \Omega} \left(
\bbox{\nabla} \varphi\right)^2\, d^3{\bf r},
\end{equation}
so that $\delta U$ can be cast in the form
\begin{equation}
\delta U = \int_{\Omega} \varphi \,\delta q \, d^3{\bf r} + 
\int_{\delta \Omega} q \, \varphi  d^3{\bf r} -
\frac{\varepsilon}{8\pi} \int_{\delta \Omega} \left(
\bbox{\nabla} \varphi\right)^2\, d^3{\bf r}.
\label{eq:deltau}
\end{equation}
From eq. (\ref{eq:s}) and relation (\ref{eq:boltzm}), the change
in entropy reads:
\begin{eqnarray}
\delta S &=& k_{_{B}} \sum_{\alpha = +,-} \int_{\delta \Omega} 
\rho^{\alpha}({\bf r})\, d^3{\bf r} +\int_{\delta \Omega} 
\varphi\, q \, d^3{\bf r}
+\int_{\Omega} \varphi \,\delta q \, d^3{\bf r}
\nonumber\\
&&-k_{_{B}} \sum_{\alpha = +,-} 
\ln\left(\rho_{_{0}}^\alpha \Lambda_\alpha^3  \right)\left[
\int_{\Omega} \delta \rho^\alpha \, d^3{\bf r} + \int_{\delta \Omega}
\rho^\alpha\, d^3{\bf r} \right]
\end{eqnarray}
where it was remembered that 
$$
\delta q_{_{\cal P}}({\bf r}) = 0 \qquad \hbox{and} \qquad 
\int_{\delta \Omega} q_{_{\cal P}}({\bf r})\, \varphi({\bf r}) \, 
d^3{\bf r} = 0.
$$
Since $\delta N^+ = \delta N^-=0$, the expression for the entropy 
simplifies to:
\begin{equation}
\delta S = k_{_{B}}  \sum_{\alpha = +,-} \int_{\delta \Omega} 
\rho^{\alpha}({\bf r})\, d^3{\bf r} +\int_{\delta \Omega} 
\varphi\, q \, d^3{\bf r}
+\int_{\Omega} \varphi \,\delta q \, d^3{\bf r}.
\label{eq:deltas}
\end{equation}
Upon substitution of eqs. (\ref{eq:deltau}) and (\ref{eq:deltas})
into $\delta F = \delta U - T\delta S$, we obtain
eq. (\ref{eq:dFshape})
\begin{eqnarray}
\delta F &=& -k_{_{B}} T \sum_{\alpha = +,-} \int_{\delta \Omega} \,
\rho^{\alpha}({\bf r}) \, d^3{\bf r} \,- \, \frac{\varepsilon}{8\pi} 
\int_{\delta \Omega} \, \left[\bbox{\nabla} \varphi\right]^2 \, d^3{\bf r}
\nonumber \\
&=& -\int_{\delta \Omega} \Pi({\bf r})\, d^3{\bf r},
\end{eqnarray}
where the following local osmotic pressure was introduced
\begin{equation}
\Pi({\bf r}) = k_{_{B}} T \sum_{\alpha = +,-} 
\rho^{\alpha}({\bf r}) + \frac{\varepsilon}{8\pi} 
\left(\bbox{\nabla} \varphi\right)^2.
\end{equation}
Near the surface $\Sigma$, the relation between $\Pi({\bf r})$ and
the pressure tensor (\ref{eq:ptenseur}) is
\begin{equation}
\Pi({\bf r}) = \bbox{n}_{_{\Sigma}}\,\cdot\!
\stackrel{\longleftrightarrow}{\bbox{\Pi}}\!
\cdot \,\bbox{n}_{_{\Sigma}},
\end{equation}
so that the osmotic pressure $\Pi$ may be  defined as the average
of $\Pi({\bf r})$ over the surface $\Sigma$ (cf. eq. (\ref{eq:pression})).

The variation $\delta F$ can be partitioned into kinetic
and electrostatic contributions
\begin{mathletters}
\begin{eqnarray}
&&\delta F^{\hbox{\scriptsize kin.}} 
= -k_{_{B}} T \sum_{\alpha = +,-} \int_{\delta \Omega} \,
\rho^{\alpha}({\bf r}) \, d^3{\bf r} 
\label{eq:dFpartitiona} \\
&&\delta F^{\hbox{\scriptsize el.}} =
- \, \frac{\varepsilon}{8\pi} 
\int_{\delta \Omega} \, \left[\bbox{\nabla} \varphi\right]^2 \, d^3{\bf r}.
\label{eq:dFpartitionb}
\end{eqnarray}
\end{mathletters}
Within linearized PB theory, we shall finally prove that, when the shape of the cell is modified at constant total volume, the resulting variation 
of the free energy is proportional to the total quadrupole moment
of the charge distribution inside the cell.

With the chosen boundary conditions, the total quadrupole defined
from eq. (\ref{eq:quadru}) is
\begin{equation}
Q_{zz}^{\hbox{\scriptsize tot.}} = \frac{1}{2}\oint_{\Sigma}
\varphi({\bf r}) \, \bbox{\nabla}\left(2z^2-x^2-y^2\right)\cdot
\bbox{d S}. 
\end{equation}
Within linearized PB theory, 
\begin{equation}
\delta F^{\hbox{\scriptsize kin.}} = \beta e \, \left(
\rho_{_{0}}^+-\rho_{_{0}}^-\right) \int_{\delta \Omega}
\varphi({\bf r}) \, d^3{\bf r}.
\end{equation}
Consider the case of a cylindrical WS cell. Taking into account
the conservation of the overall volume, it is straightforward to show that  \begin{equation}
H \frac{\delta F^{\hbox{\scriptsize kin.}}}{\delta H}
\biggl|_{T,\Omega, N^\alpha}
 = \frac{1}{2}\,
\gamma_{\scriptscriptstyle 0}\, \kappa_{_{D}}^2 
Q_{zz}^{\hbox{\scriptsize tot.}}
\label{eq:Fcinquadru}
\end{equation}
A similar result holds for a parallelepipedic cell.

\newpage

\begin{center}
{\large FIGURES}
\end{center}

Fig 1. The prismatic cells considered in this paper. Fig 1a shows a 
circular platelet
in a cylindrical cell, whereas Fig 1b represents the parallelepipedic
cell containing either a circular or a square platelet lying in the 
dashed region $S'$. 

\bigskip

Fig. 2. Variations of the free energy $F$, the osmotic pressure $\Pi$
and the disjoining pressure $\Pi_d$ with the aspect ratio (upper
curves, part a)). For illustrative purposes,
the free energy has been shifted by an arbitrary constant
along the
vertical axis. 
Also shown is the total normalized quadrupole
(lower part, Fig 2b)). The kinetic and electrostatic parts of the
free energy variation are defined in the appendix
(cf eq. (\ref{eq:dFpartitiona}) and (\ref{eq:dFpartitionb})).
In order to check the validity of 
eq. (\ref{eq:Fcinquadru}), we choose
$\alpha^{-1} = \gamma_{\scriptscriptstyle 0}\, \kappa_{_{D}}^2 
Q_{zz}^{\hbox{\scriptsize disc}}/2$. The normalization pressure 
$\Pi_0$ is defined with the macroscopic concentrations of 
co- and counterions inside the cell: $\Pi_0 = k_{_{B}} T \,(\rho_{_{0}}^+ + 
\rho_{_{0}}^-)$.

\bigskip

Fig. 3. Variation of the optimal aspect ratio minimizing the free energy
with clay concentration. The data shown correspond to a circular platelet
in a cylindrical cell,
with $n'_s= 10^{-3}\,$M.

\bigskip

Fig. 4. Equipotential lines in the plane 
$(\kappa_{_{D}}r,\kappa_{_{D}}z)$, with arithmetic spacing between the
isopotentials
($\beta e \, \Delta \varphi = 0.5$ between two succesive curves).
Here, $Z=100$, $n=10^{-5}\,$M and $n_s=10^{-3}\,$M. 
For these parameters, the optimal aspect ratio is 
$h/r_{\scriptscriptstyle0} \simeq 2.0$. The reduced radii
of the disc and cylinder are respectively
$\kappa_{_{D}} r_{\scriptscriptstyle 0} \simeq 1.59$ and
$\kappa_{_{D}} R \simeq 4.14$. The summation in eq. (\ref{eq:seriecyl})
was truncated after $n_{max} = 50$.

\bigskip

Fig. 5a. Clay concentration dependence of the osmotic and disjoining
pressures, evaluated at the optimal aspect ratio minimizing the 
free energy (where $\Pi = \Pi_d$). The salt concentration in the
reservoir is held constant ($n'_s = 10^{-3}\,$M) and defines
the normalization pressure 
$\Pi^{\hbox{\scriptsize reservoir}} = 2 k_{_{B}}T\, n'_s$.

\bigskip

Fig. 5b. Variation of the osmotic and disjoining pressures with the
salt concentration in the reservoir ($n'_s$). For $n'_s \to 0$,
$\Pi/\Pi^{\hbox{\scriptsize res.}}$ diverges like $1/n'_s$,
while $\Pi \to \hbox{Cst}$.

\bigskip

Fig. 6a. The characteristic double-layer thickness (see eq. 
(\ref{eq:lambda}))
as a function of
clay concentration, for the optimal aspect ratio minimizing $F$. 
The salt 
concentration in the reservoir is $n'_s = 10^{-3}\,$M. 

\bigskip

Fig 6.b. Variation of the reduced double-layer thickness 
with salt concentration
for $n=5.10^{-5}\,$M.

\bigskip

Fig. 7. Free energy and total normalized quadrupole {\it vs.} 
$h/r_{\scriptscriptstyle 0}$ for discs in cylindrical
and parallelepipedic WS cell 
($r_{\scriptscriptstyle 0}=125\,$\AA); same quantity as a function of
$H/l_{\scriptscriptstyle 0} = 2 h/l_{\scriptscriptstyle 0}$,
for a square
platelet of identical area 
and surface charge
in a parallelepipedic cell ($l_{\scriptscriptstyle 0}=221\,$\AA). 
The vertical lines
emphasize the correlation between the minimum of the free
energy and the vanishing of the total quadrupole.

\bigskip

Fig. 8. Free energy, osmotic and disjoining pressures as
functions of the aspect ratio 
$H/l_{\scriptscriptstyle 0}$,
for a square
platelet in a parallelepipedic cell. The data correspond to
$n = 10^{-4}\,$M, $n'_s=10^{-3}\,$M, $Z=100$ and 
$l_{\scriptscriptstyle 0} = 250\,$\AA.

\bigskip

Fig. 9. Variation of the force $F_z$ with the distance $d$
between the centres of co-axial discs
($F_{\scriptscriptstyle 0}=4 \pi^2 r_{\scriptscriptstyle 0}^2 \sigma^2
/\varepsilon$). The four curves correspond to different salt concentrations.

\bigskip

Fig. 10. Limits of validity for the linearized approximation of PB
theory, for a disc of radius $r_{\scriptscriptstyle 0}$
in the infinite dilution limit. The dashed line corresponds to the
case of an infinite plane (in which case $r_{\scriptscriptstyle 0}$
is an arbitrary normalization length). 
$\lambda_{_{D}}$ denotes the Debye length
and $b$ is the Gouy length.

\end{document}